\documentclass[preprint,longbibliography,showpacs,preprintnumbers,amssymb,superscriptaddress,aps,prd,nofootinbib,11pt]{revtex4-1}
\usepackage[left=2.5cm,right=2.5cm,top=2.5cm,bottom=2.5cm]{geometry}

\selectfont

\usepackage[utf8]{inputenc}
\usepackage[]{graphicx}
\usepackage{url}
\usepackage[usenames]{color}
\usepackage[usenames,dvipsnames,svgnames,table]{xcolor}
\usepackage{multirow}
\usepackage{ulem}
\usepackage[scr=boondoxo,scrscaled=1.05]{mathalfa}%script fonts
\usepackage{tensor}
\usepackage{booktabs}
\usepackage{amsmath}
\usepackage{subfigure}

\newcommand{\beq}{\begin{equation}}
\newcommand{\eeq}{\end{equation}}

\newcommand{\uin}{u^{\text{in}}_{\lambda \omega}}
\newcommand{\uup}{u^{\text{up}}_{\lambda \omega}}
\newcommand{\um}{u^{-}_{\lambda \omega}}
\newcommand{\up}{u^{+}_{\lambda \omega}}

\usepackage[]{hyperref}
\hypersetup
	{
		colorlinks,%
		citecolor=blue,%
		linkcolor=red,%
		urlcolor=blue,%
	}

\begin{document}

\title{Dynamical system approach to the spectral (in)stability of black holes under localised potential perturbations}

\author{Theo Torres}\email{theo.torres@ube.fr}
\affiliation{Universit\'e Bourgogne Europe, CNRS, Laboratoire Interdisciplinaire Carnot de Bourgogne ICB UMR 6303, 21000 Dijon, France}

%\affiliation{Department of Physics,King's College London,The Strand,London WC2R 2LS,United Kingdom}

\author{Sam R.~Dolan}
\affiliation{
Consortium for Fundamental Physics,
School of Mathematics and Statistics,
University of Sheffield,
Hicks Building,
Hounsfield Road,
Sheffield S3 7RH, 
United Kingdom}

\begin{abstract}
The aim of this work is to improve understanding of the resonant spectra of black holes under perturbations arising from e.g.~compact objects or accretion disks in their vicinity. It is known that adding a weak perturbation to the radial potential can strongly disrupt the spectrum of quasinormal modes and Regge poles of a black hole spacetime. Here we examine the effect of (weak or strong) localised delta-function perturbations on the resonant spectra of spherically-symmetric systems, to address fundamental questions around linear and non-linear spectral stability. We examine two cases: the Nariai spacetime with a Poschl-Teller potential and the Schwarzschild spacetime. We show that, in either case, the spectrum deforms in a smooth and continuous manner as the position and strength of the perturbation is varied. As the strength of the perturbation is increased, resonances migrate along trajectories in the complex plane which ultimately tend towards attracting points determined by a hard-wall scenario. However, for weak perturbations the trajectory near the unperturbed resonance is typically strongly influenced by a set of repelling points which, for perturbations far from the system, lie very close to the unperturbed resonances; hence there arises a non-linear instability (i.e.~the failure of a linearised approximation). Taking a dynamical systems perspective, the sets of attracting and repelling spectral points follow their own trajectories as the position of the perturbation is varied, and these are tracked and understood.
\end{abstract}

\maketitle

\section{Introduction}

Resonances of black hole geometries were first identified more than fifty years ago in studies of time-dependent gravitational-wave scattering \cite{Vishveshwara:1970}. The prediction that black holes possess characteristic complex frequencies—encoding both oscillation and decay—was spectacularly confirmed by the first direct detection of gravitational waves in 2016 \cite{LIGOScientific:2016aoc} and by many subsequent observations~\cite{LIGOScientific:2021djp}. These detections established that, following a merger, the remnant black hole undergoes a rapid “ringdown” phase dominated by a small number of quasinormal modes (QNMs), most prominently the fundamental quadrupolar mode ($\ell =2$, $n=0$). Each QNM is characterised by a complex frequency whose real part determines the oscillation frequency and whose imaginary part sets the damping timescale. As a result, black hole quasinormal mode spectra have become a central object of study in gravitational physics \cite{Leaver:1985ax,Leaver:1986gd}. 

Alongside QNMs, a second class of resonances—the Regge poles—also plays an important role in black hole scattering theory \cite{Decanini:2002ha}. Within the framework of complex angular momentum (CAM) analysis, Regge poles encode interference phenomena associated with unstable photon orbits, including glory and spiral scattering effects \cite{Folacci:2019cmc,Folacci:2019vtt}, and regular oscillations of the absorption cross section \cite{Decanini:2011xi,Decanini:2011xw}. Since the Quasinormal Mode (QNM) and Regge pole (RP) spectra have formally-similar mathematical definitions (see Sec.~\ref{subsec:resonances}), herein we shall refer to both as black hole resonances.

A compelling geometric interpretation of black hole resonances was proposed by Goebel \cite{Goebel:1972}, who associated them with a congruence of null rays orbiting the black hole near its light-ring. This intuitive view is supported by time-domain simulations in black hole perturbation theory, in which the `ringing' is clearly generated in the vicinity of the peak of the potential barrier, near $r=3M$. Making this more precise, in Ref.~\cite{Cardoso:2008bp} it was shown that, in the geometric-optics limit ($\ell \gg 1$, $\ell \gtrsim n$), the real and imaginary part of the quasinormal frequencies (QNFs) of a stationary, spherically symmetric and asymptotically flat black hole are determined by the orbital frequency and expansion rate (Lyapunov exponent) of this congruence. Further supporting the geometric interpretation, for Kerr black holes the asymptotic spectrum of QNFs is determined by the local properties of the family of photon orbits \cite{Dolan:2010wr,Yang:2012he,Fransen:2023eqj}. Following from the close relationship between QNM and RPs, a similar interpretation holds in the CAM approach \cite{Decanini:2002ha,Dolan:2009nk}.

The geometric interpretation is limited, however, for precisely the same reason that it is appealing: it is local, associating resonances with wave-generation near the light ring. However, QNMs (and RPs) are global, in the sense that they are formally defined in the frequency domain in terms of outgoing boundary conditions in two asymptotic regions (typically, the horizon and spatial infinity). They are eigenvalues of a non-Hermitian operator and with the non-Hermiticity comes a number of remarkable features~\cite{ashida2020non}. Amongst them is the phenomena of spectral instability: slightly modifying the boundary conditions or perturbing the potential will drastically modify the spectrum. The question of precisely how the resonance spectrum is modified by small perturbations is the topic of pioneering studies \cite{Nollert:1996rf,Nollert:1998ys, Leung:1997was,Leung:1999iq,Barausse:2014tra} and a flurry of recent work \cite{Jaramillo:2020tuu,Jaramillo:2021tmt,Berti:2021ijo,Berti:2022xfj,Cheung:2021bol,Konoplya:2022iyn,Torres:2023nqg,rosato2024ringdown,yang2024spectral,Syu:2025rex}. 

We emphasise that the strong deformation of the resonant spectrum can be triggered by two fundamentally distinct mechanisms: i) by global high-frequency/random perturbation~\cite{Jaramillo:2020tuu} and ii) by local modification of the potential~\cite{Cheung:2021bol}. In the first scenario, the norm of the perturbation is precisely defined and controlled and allows for an unambiguous identification of the spectral instability effect. The second scenario is mathematically more vague but physically more intuitive. 
%We expect the environment of the black hole to generate a local perturbation with a characteristic size which is `small' compared to the unperturbed potential. 
It is this second scenario, usually refered to as the `Elephant and Flea', that we investigate in the paper.
In general, the QNM spectrum is found to be highly sensitive~\cite{Barausse:2014tra}, and subject to the `Elephant and Flea' phenomenon~\cite{Cheung:2021bol}: a tiny perturbation in the potential placed very far from the system can entirely reconfigure the overtones of the QNM spectrum in the complex plane. The higher overtones of the Regge pole spectrum are also found to be highly sensitive \cite{Torres:2023nqg}, though to a lesser degree which we quantify here. 

This sensitivity raises questions about the robustness of the black hole spectroscopy program which aims at recovering black hole parameters from the emitted ringdown signal. While signs of the perturbed spectrum where found in the emitted GW signals of a perturbed BH~\cite{Jaramillo:2021tmt},
it has also been shown that, observable quantities linked to the QNMs and RPs remain stable~\cite{Torres:2023nqg,rosato2024ringdown}.

The aim of this work is to clarify the nature of this spectral sensitivity by studying the deformation of black hole resonance spectra under localised potential perturbations. We focus on idealised delta-function perturbations, which allow us to isolate fundamental mechanisms governing linear and non-linear spectral stability. By adopting a dynamical-systems perspective, we show that resonances migrate along well-defined trajectories in the complex plane, influenced by sets of attracting and repelling points whose structure explains both the apparent instability of perturbation theory and the smooth global deformation of the spectrum.
To develop intuition, we analyse in depth the problem in the Nariai spacetime with a Pöschl–Teller potential, before applying our framework to the Schwarzschild black hole.
A key advantage of studying the Nariai problem is that the radial functions, and the Wronskian, are available in closed form in terms of special functions. This allows us to illuminate certain features of the Schwarzschild case.

The paper is organised as follows. In Sec.~\ref{sec:formulation} we introduce a general framework for studying resonances under potential perturbations. We formulate the resonance condition in terms of the Wronskian of appropriately defined radial solutions, derive series expansions for perturbed resonance frequencies, and introduce quantitative measures of linear and non-linear spectral stability. We also reinterpret the deformation of the resonance spectrum as a dynamical flow in the complex plane, identifying the roles of attractors and repellers. In Sec.~\ref{sec:Nariai} we apply this formalism to the Nariai spacetime, and perform for a complete analysis of quasinormal modes and Regge poles under perturbation. Section~\ref{sec:Schwarzschild} extends the analysis to the physically relevant case of the Schwarzschild spacetime, where we examine the migration, and stability of both QNM and RP spectra. Finally, in Sec.~\ref{sec:conclusions} we summarise our results and discuss their implications for the interpretation of spectral instability.

\section{Formulation\label{sec:formulation}}

 \subsection{Resonances under perturbation\label{subsec:resonances}}

The main motivation of this paper is the investigation of the spectral instability of black hole spacetimes. However, we begin our discussion by considering a generic 3+1 (2+1) dimensional system with spherical (cylindrical) symmetry which reduces, after separation of variables, to solving a radial equation of the form
\beq
\left\{ \frac{d^2}{dx^2} + V_{\lambda \omega} (x) + \epsilon \, \delta V (x) \right\} u^{(\epsilon)} (x) = 0 . \label{eq:radial-equation}
\eeq
where $\omega$ and $\lambda$ are parameters representing the wave frequency and the angular momentum, respectively.
For example, in the case of a scalar field on Schwarzschild spacetime, this reduces to the Regge-Wheeler equation Eq.~\eqref{eq:RW1} with the potential given in Eq.~\eqref{eq:rw} and $x$ is the tortoise coordinate. The term $\epsilon \delta V(x)$ represents a localised perturbation to the potential that we shall assume to be of compact support  (see Sec.~\ref{subsec:potentials} for specifics).

Generically, for such an equation there exist homogeneous solutions $\uin(x)$ and $\uup(x)$ that satisfy the physically-motivated boundary conditions on the left-hand side and the right-hand side of the domain in $x$. In the following, the radial functions are analytically continued to allow the parameters $\lambda$ and $\omega$ to take complex values.  

From those homogeneous solutions, one can construct the Green's function, $G(x,x';\lambda,\omega)$, which is commonly used to study the system’s response to external perturbations. Explicitly the Green's function is given by:
\begin{equation}
    G(x,x';\lambda,\omega) = \frac{1}{W}\begin{cases}
        u^{\rm in}_{\lambda\omega}(x)u^{\rm up}_{\lambda\omega}(x')\quad \text{if} \quad x\leq x', \\
        u^{\rm up}_{\lambda\omega}(x)u^{\rm in}_{\lambda\omega}(x')\quad \text{if} \quad x'\leq x.
    \end{cases}
\end{equation}
In the above, $W$ is the Wronskian, defined in the standard way,
\beq
W\left( \uin ,\uup \right) \equiv \uin \frac{d \uup}{dx} - \uup \frac{d \uin}{dx} .
\eeq
A resonance occurs where there exists a mode that satisfies both physical boundary conditions. Equivalently, the two radial solutions $IN$ and $UP$ are no longer linearly independent, and so
\beq
W\left( \uin ,\uup \right) = 0 , \label{eq:resonance-condition}
\eeq
From this definition, we see that the resonances coincides with the poles of the Green's function~\cite{Leaver:1986gd,Berti:2006wq}.
For real $\lambda$, the condition $W = 0$ determines a (complex) spectrum of quasinormal mode frequencies (QNFs), $\mathcal{Q}_{\lambda} (\epsilon) = \{ \omega_n \}$. For real $\omega$, the condition $W = 0$ determines a (complex) spectrum of Regge pole values (RPVs), $\mathcal{R}_{\omega} (\epsilon) \equiv \{ \lambda_{n} \}$ where $\lambda \equiv \ell + 1/2$. Here $n \in \{0, 1, 2, \ldots\}$ is the overtone number. In this work we study the deformation of these spectra as the perturbation strength $|\epsilon|$ is increased from zero. 

For a perturbation with no support outside $x \in [a,b]$, the application of regular perturbation theory (see Appendix \ref{app:perturbation-theory}) yields, for $x>b$,
\begin{subequations}
\begin{align}
\uin(x > b) &= \um(x) + \frac{\epsilon}{W(\um,\up)} \left( c^-_{\lambda \omega} \um(x) - c^+_{\lambda \omega} \up(x) \right) + O(\epsilon^2) , \\
\uup(x > b) &= \up(x) ,
\end{align}
\label{eq:u-general}
\end{subequations}
up to a multiplicative scaling, where $\um \equiv \uin(\epsilon=0)$ and $\up \equiv \uup(\epsilon=0)$ are the corresponding functions for the unperturbed potential, and
\beq
c^{\pm}_{\lambda \omega} \equiv \int_{a}^{b} \delta V(x') u^{\mp}(x') \um(x') dx' .  \label{eq:cco}
\eeq
Hence the Wronskian is
\beq
W(\uin, \uup) = W_0 + \epsilon \, c^-_{\lambda \omega} + O(\epsilon^2). \label{eq:Wronsk1}
\eeq
where $W_0$ is the Wronskian for the unperturbed case,
\beq
W_0 \equiv W(\um, \up) .  \label{eq:W0}
\eeq

In the following we consider the special case of a Dirac delta-function in the perturbed potential: $\delta V(x) = \delta(x - x_0)$.  
Matching at $x=x_0$ then yields
\begin{align}
\uin(x) = \um(x) +  \Theta(x-x_0) \frac{\epsilon \, \um(x_0) }{W(\um,\up)} \left( \up(x_0) \um(x) - \um(x_0) \up(x) \right) 
\end{align}
where $\Theta(x - x_0)$ is the Heaviside step function. The corresponding solution for $\uup(x)$ follows by symmetry. Consequently, the Wronskian is 
\beq
W(\uin, \uup) = W_0 + \epsilon \, \um(x_0) \up(x_0) . \label{eq:Wronsk2}
\eeq
Note that Eq.~(\ref{eq:Wronsk2}) is consistent with Eq.~(\ref{eq:Wronsk1}) and moreover it is exact, i.e., $O(\epsilon^2)$ corrections are absent. Hence the resonance condition (\ref{eq:resonance-condition}) reduces to the equation
\begin{subequations}
\beq
F(\lambda, \omega) \, + \, \epsilon = 0 
\eeq
where 
\beq
F(\lambda, \omega) \equiv \frac{W(\um, \up)}{\um(x_0) \up(x_0)} . 
\eeq
\label{eq:delta-resonance}
\end{subequations}
Through Eq.~(\ref{eq:delta-resonance}), we can now deduce the spectrum of the perturbed potential entirely from the properties of the radial functions of the \textit{unperturbed} potential, $\um(x_0)$ and $\up(x_0)$.

 \subsection{Series expansion and the excitation factors\label{subsec:series}}
For sufficiently small $\epsilon$, one would expect that a Taylor-series expansion derived from Eq.~(\ref{eq:delta-resonance}) would provide a good approximation, viz.,
\beq
z_n(\epsilon) = z_n(0) - \alpha_{1} \epsilon - \alpha_{2} \epsilon^2 + \ldots \label{eq:series}
\eeq
where
\beq
\alpha_1 = \frac{1}{F'_0} , \quad \quad \alpha_2 = \frac{F''_0}{2 (F'_0)^3} ,
\eeq
and here $z$ represents either $\omega$ (for QNMs) or $\lambda$ (for RPs), and $F_0' \equiv \frac{dF}{dz}(z_n(0))$,  $F_0'' \equiv \frac{d^2F}{dz^2}(z_n(0))$, etc. 

In the QNM case, there is a close relationship between the linear coefficient in Eq.~(\ref{eq:series}), that is,
\beq
\alpha_{\lambda n}(x_0) \equiv \frac{u^{-}_{\lambda \omega_n} (x_0) u^{+}_{\lambda \omega_n}(x_0)}{ \left. \frac{\partial W}{\partial \omega} \right|_{\omega=\omega_n} } , \label{eq:linearised}
\eeq
and the so-called QNM excitation factors $\mathcal{B}_{\lambda n}$. The latter arise in the context of the initial-value problem, in addressing the question of how much each QNM is excited by given initial data \cite{Leaver:1986gd, Sun:1988tz, Sun:1990pi, Andersson:1996cm, Nollert:1998ys, Berti:2006wq,Dolan:2011fh}. It is straightforward to show that the excitation factors, defined in e.g. Eq.~(3.8) of Ref.~\cite{Berti:2006wq}, are
\beq
\mathcal{B}_{\lambda n} = \lim_{x_0 \rightarrow \infty} i \alpha_{\lambda n} (x_0) e^{-2 i \omega_n x_0} .  \label{eq:ef}
\eeq
This limit is well-defined. This indicates that, for $\text{Im} ( \omega_n) < 0$, the linear coefficient $\alpha_{\lambda n}(x_0)$ in Eq.~(\ref{eq:linearised}) grows exponentially without bound as $x_0 \rightarrow \infty$ (for $\epsilon > 0$). This is one manifestation of the elephant and the flea phenomenon: a tiny perturbation in the potential placed very far from the system can destabilise the QNM spectrum; and this issue grows with the overtone number $n$. A closer consideration of stability is  warranted. 

% In the Regge pole case,  the frequency is real, and so the linear coefficient should not suffer the issue (or least, not the same extent). 

 \subsection{Stability and sensitivity\label{subsec:stability}}
%A discussion of the stability of the spectrum can be approached from a number of directions. Here we propose some working definitions to make the discussion in the following sections more precise. 

For a given $x_0$, it is natural to call a mode \textit{linearly unstable} if the linear coefficient in Eq.~(\ref{eq:series}) is large, $|\alpha_{1}| \gg 1$. Relatedly, we may define the perturbation strength $\epsilon_{\text{lin}}$ at which the shift in the spectrum under perturbation becomes of order unity; this is simply
\beq
\epsilon_{\text{lin}} \equiv \left| F_0' \right| .  \label{eq:epsilon-lin}
\eeq
From the discussion above, we expect quasinormal modes (particular those with large $n$) to be linearly unstable in general for large $x_0$. We shall say that a quasinormal mode is \textit{anomalously unstable} if $|\beta_{\lambda n}| \gg 1$, where, motivated by Eq.~(\ref{eq:ef}), we define
\beq
\beta_{\lambda n}(x_0) \equiv \alpha_{1}(x_0) e^{-2 i \omega_n x_0} . \label{eq:beta-def-qnm}
\eeq

A second notion of stability is linked to the validity of the truncated Taylor-series itself. We define $\epsilon_{\text{nonlin}}$ as the perturbation strength for which the quadratic term in Eq.~(\ref{eq:series}) is comparable to the linear term; more precisely,
\beq
\epsilon_{\text{nonlin}} \equiv \left| \frac{\alpha_1}{\alpha_2} \right| = \left| \frac{2 (F'_0)^2}{F''_0} \right| . \label{eq:epsilon-nonlin}
\eeq
We shall call a mode \textit{nonlinearly unstable} if $\epsilon_{\text{nonlin}} \ll 1$. 
For $|\epsilon| \gtrsim \epsilon_{\text{nonlin}}$, one would anticipate that the truncated series Eq.~(\ref{eq:series}) is not accurate for determining the perturbed mode frequencies. Moreover, the Taylor-series expansion itself is only valid within a radius of convergence, linked to the distance to the nearest pole of $F(z)$ in the complex plane. For $|\epsilon| \gtrsim \epsilon_{\text{nonlin}}$, the global properties of $F(z)$ play an important role. 

%It is naturally to say that a mode is non-linearly unstable if the quadratic term in the expansion is comparable to the linear term. 
%\beq
%\beta = \frac{1}{2} \frac{ \left. \frac{\partial^2 f}{\partial \omega^2} \right|_{\omega_n} }{ \left. \frac{\partial f}{\partial \omega} \right|_{\omega_n} }
%\eeq
%where $f$ is defined in Eq.~(\ref{eq:delta-resonance}).

 \subsection{Flows, attractors and repellers\label{subsec:flows}}
 
Under the introduction of a perturbation, the QNFs and RPVs typically migrate along trajectories in the complex plane. 
To describe a continuous deformation, the equation (\ref{eq:delta-resonance}) can be converted into a first-order autonomous ordinary differential equation,
\beq
\frac{d z}{d \epsilon} = g(z), \quad \quad g(z) \equiv - \frac{1}{F'(z)} , \quad \quad z(0) = z_n. \label{eq:ode}
\eeq
For QNMs $z = \omega$ and $\lambda$ is held constant; for RPs $z = \lambda$ and $\omega$ is held constant. Here $F'(z) = \partial_z F$ and $z_n$ is an initial value in the unperturbed spectrum. The integral curves $z(\epsilon)$ represent trajectories of the QNFs (or RPVs) in the complex plane under perturbation. Taking a dynamical systems perspective, we now consider the poles and zeros of $g(z)$, the right-hand side of the ODE in Eq.~(\ref{eq:ode}).

The simple zeros of $g(z)$ define \textit{fixed points}.
In the vicinity of a fixed point, the trajectory can be written $z(\epsilon) = \bar{z} + \delta{z}(\epsilon)$, where $g(\bar{z}) = 0$. Taking a series expansion of the ODE~\eqref{eq:ode} leads to the solution $|\delta{z}| \propto \exp({\rm Re}[g’(\bar{z})] \epsilon)$, so for ${\rm Re}[g’(\bar{z})] < 0$ (${\rm Re}[g’(\bar{z})] > 0$) the point is an attractor for $\epsilon \rightarrow \infty$ ($\epsilon \rightarrow -\infty$). 
%For the special case ${\rm Re}[g’(\bar{z})] = 0$, the fixed point is a centre. 
We shall call fixed points with ${\rm Re}[g’(\bar{z})] \neq 0$ \textit{attracting points}.
%limit points of trajectories $z(\epsilon)$ as $\epsilon \rightarrow \infty$ (or $\epsilon \rightarrow -\infty$). 
The simple zeros of $g(z)$ correspond to simple poles of
$F'(z)$. In turn, these poles are typically associated with simple zeroes in either $u^+_{z}(x_0)$ or $u^-_{z}(x_0)$. In physical terms, a zero boundary condition for the mode function at $x = x_0$ is necessary if there is a `hard wall' in the potential at $x=x_0$; so it is not surprising to find the zeros of $u^+_{z}(x_0)$ or $u^-_{z}(x_0)$ correspond to attractors as the strength of the delta-function perturbation increases.

The simple poles of $g(z)$ define \textit{repelling points}. Near a repelling point $z=z_r$, 
\beq
F(z) + \epsilon \; \approx \; i \, \text{Im} \, F(z_r) + (z - z_r)^2 F''(z_r) + \Delta \epsilon , \label{eq:f-repelling}
\eeq
where $\Delta \epsilon = \epsilon + \text{Re} \, F(z_r)$. As $\Delta \epsilon$ changes sign, the resonance trajectory, defined by the root of Eq.~(\ref{eq:f-repelling}), will rapidly change its direction in the complex plane by $90^\circ$. 

For the special case $\text{Im} \, F(z_r) = 0$, the term `repelling point' is a misnomer; in the complex plane, pairs of trajectories approach the repelling point along antiparallel directions, pass through $z=z_r$, and then depart in orthogonal antiparallel directions. Hence, we will refer to such points as \textit{junction points}.

  \subsection{Schwarzschild spacetime and the comparison problem\label{subsec:potentials}}

Here we define the specific potentials that we shall investigate in the remainder of the article.

 A scalar field $\Phi(x)$ on Schwarzschild spacetime satisfying $\Box \Phi = 0$ admits a full separation of variables
 \beq
 \Phi(x) = \frac{1}{r} \sum_{\ell m} \int d \omega \,  c_{\ell m \omega} \, u_{\ell \omega} (r) Y_{\ell m}(\theta) e^{- i \omega t + i m \phi} 
 \eeq
 and the radial functions $u_{\ell \omega} (r)$ satisfy the Regge-Wheeler equation
 \begin{align}
\left\{ \frac{d^2}{dx^2} + \omega^2 - V_{\ell}(r) \right\} u_{\ell \omega} = 0 ,\label{eq:RW1} \\ 
\quad \quad V_\ell(r) = f \left( \frac{\ell (\ell+1)}{r^2} + \frac{2M (1-s^2)}{r^3} \right), \label{eq:rw}
 \end{align}
 where $f(r) = 1 - 2M/r$. This equation is satisfied by electromagnetic ($s=1$) and linearised-gravitational ($s=2$) fields as well as the scalar field ($s=0$). The tortoise coordinate $x$ is defined by $\tfrac{dr}{dx} = f(r)$; we choose the constant of integration so that the light-ring at $r=3M$ maps to $x=0$ (i.e.~$x = r + 2M \ln(r / 2M - 1) + M (2 \ln 2 - 3)$).

 In the following, we shall also consider a comparison problem for which closed-form solutions are available, namely the Nariai spacetime ($dS_2 \times S_2$) \cite{Nariai:1951} with line element
\beq
ds^2 = -h(\rho) dt^2 + h^{-1}(\rho) d \rho^2 + d \Omega_2^2, 
\eeq
with $h(\rho) = 1- \rho^2$ and $\rho \in (-1,1)$ \cite{Casals:2009zh}. The (non-minimally-coupled) Klein-Gordon equation $(\Box + \tfrac{1}{8} R) \Phi = 0$ admits a separation of variables $\Phi = u_{\lambda \hat{\omega}}(x) e^{-i \hat{\omega} t} Y_{\ell m}(\theta,\phi)$ leading to the radial equation \cite{Casals:2009zh}
\beq
\left\{ \frac{d^2}{d\hat{x}^2} + \hat{\omega}^2 - \frac{\lambda^2 + 1/4}{\cosh^2 \hat{x}} \right\} u_{\lambda \hat{\omega}} = 0, \label{eq:nariai-radial}
\eeq
where $\rho = \tanh \hat{x}$. The potential barrier is of P\"oschl-Teller type, with a single maximum at $\hat{x}=0$. For a closest match to the Schwarzschild barrier in the vicinity of the peak one should make the association $x \leftrightarrow \nu \hat{x}$, $\omega \leftrightarrow \hat\omega / \nu$, where $\nu=\sqrt{27} M$. %, $x$ is the Schwarzschild tortoise coordinate and $\omega$ is the Schwarzschild frequency.

\section{Perturbed resonances of the Nariai spacetime}\label{sec:Nariai}

The principal advantage of working with the Nariai comparison problem is that the potential has a single peak (as in the Schwarzschild case) and Eq.~(\ref{eq:nariai-radial}) admits solutions in closed form, viz.,
\beq
\up(\hat{x}) = e^{i \omega \hat{x}} {}_{2}F_{1}\left( \tfrac{1}{2} - i \lambda, \tfrac{1}{2} + i \lambda, 1 - i \hat{\omega}, \frac{1}{1 + e^{2\hat{x}}} \right) , \label{eq:uup-pt}
\eeq
with $\um(\hat{x}) \equiv \up(-\hat{x})$. The corresponding Wronskian is
\beq
W( \um, \up ) = 2 i \hat{\omega} \frac{\Gamma(1- i \omega) \Gamma( - i \omega)}{\Gamma(\tfrac{1}{2} + i (\lambda - \hat{\omega}) ) \Gamma(\tfrac{1}{2} - i (\lambda + \hat{\omega}))} \label{eq:Wronski-Nariai}
\eeq
The unperturbed resonances are determined from the condition $W( \um, \up) = 0$, and thus from the poles of the gamma functions in the denominator of Eq.~(\ref{eq:Wronski-Nariai}). The unperturbed spectra are
\begin{subequations}
\begin{align}
\mathcal{Q}_{\lambda} &= \left\{ \hat{\omega}_n \equiv \pm \lambda - i (n + \tfrac{1}{2}) \right\}, \\
\mathcal{R}_{\hat{\omega}} &= \left\{ \lambda_n \equiv \pm \left( \hat{\omega} + i (n + \tfrac{1}{2}) \right) \right\}. 
\end{align}
\end{subequations}
Henceforth we shall restrict attention to spectrum in quadrants I and IV of the complex plane (as the spectrum in quadrants II and III follows from symmetry). 

The condition for a perturbed resonance, Eq.~(\ref{eq:delta-resonance}), can be solved in closed form in certain limiting cases.

\subsection{Quasinormal mode spectrum}

\begin{figure}
 \includegraphics[width=14cm]{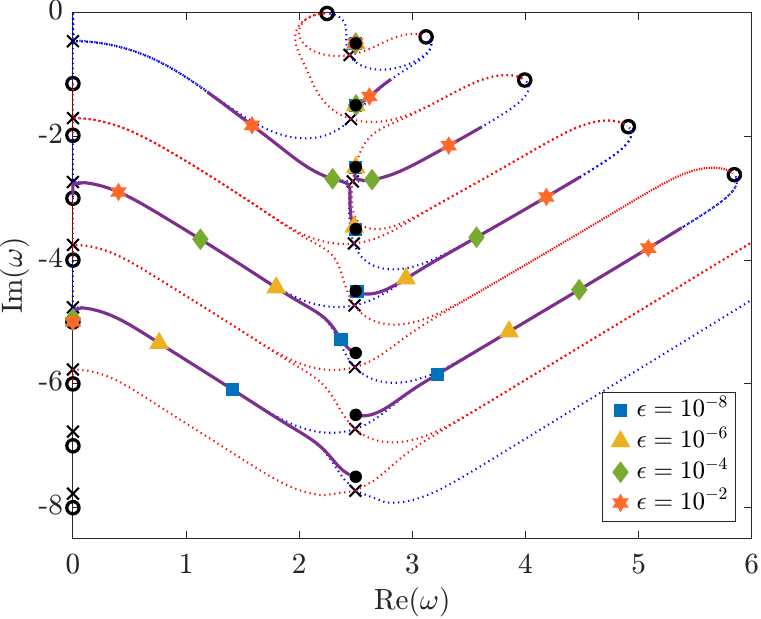} \\
\caption{
The quasinormal mode spectrum for a P\"oschl-Teller potential (\ref{eq:nariai-radial}) with a delta-perturbation at $\hat{x}_0 = 10 / \sqrt{27}$ of strength $\epsilon$ and angular momentum $\lambda = \ell + \tfrac{1}{2} = \tfrac{5}{2}$. The resonances migrate along trajectories shown as solid purple lines, ultimately towards the attracting points marked as open black circles. The unperturbed spectrum is shown with filled black circles and the perturbed spectrum for $\epsilon \in \{10^{-8}, 10^{-6}, 10^{-4}, 10^{-2} \}$ is shown with filled points (square, triangle, diamond and stars). The spectrum is symmetric about the imaginary axis. Trajectories that meet at $\text{Re} \ \omega = 0$ (from either side) then migrate along the imaginary axis to the attractors that lie (in alternating fashion) just above and below the points $\omega = -in$. The repelling points are marked with black crosses, and the integral curves of Eq.~(\ref{eq:ode}) that move away from (towards) these points as $\epsilon$ increases are shown as dotted blue (red) lines.
}
\label{fig:PT-QNMs}
\end{figure}

Figure \ref{fig:PT-QNMs} shows the first eight modes of the QNM spectrum for the Nariai spacetime (i.e.~P\"oschl-Teller potential) with $\lambda=5/2$ with a perturbation at $x_0 = 10 / \sqrt{27}$. Under perturbation, the spectrum exhibits two branches, with modes moving to the left and right in an alternating fashion. 
The high overtones are very sensitive to small perturbations: note, for example, the shift in the $n=7$ mode under a perturbation of $\epsilon = 10^{-8}$. By comparison, the fundamental mode $n=0$ is less sensitive and is least affected by the perturbation.

The QNFs migrate along trajectories that, in the $\epsilon \rightarrow \infty$ limit, terminate at attracting points corresponding to zeroes of $\um(x_0)$ or $\up(x_0)$. For $x_0>0$, the latter zeros lie on the imaginary axis. The spectrum is symmetric under reflection in the imaginary axis. Trajectories meet the imaginary axis at junction points (i.e. repelling points with ${\rm Im} \ F(z_r)=0$, see Sec.~\ref{subsec:flows}) and then migrate along it. There is one (complex) repelling point associated with each overtone, sitting somewhat below the unperturbed QNF in the complex plane. The integral curves of Eq.~(\ref{eq:ode}) that start at the repelling points (dotted lines) quickly approach the QNF trajectories.

\begin{figure}
 \includegraphics[width=10cm]{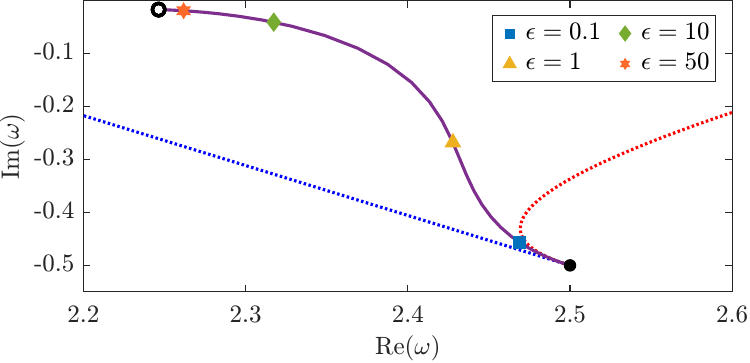} \\
\caption{
Example of the migration of the fundamental ($n=0$) quasinormal mode frequency in the perturbed Poschl-Teller potential for $\lambda_0 = 5/2$, $\hat{x}_0 = 10 / \sqrt{27}$.  The QNF migrates from the unperturbed frequency $\lambda_0 - i /2$ (filled black circle) towards the frequency at which $\um(x_0)$ is zero (open black circle) along the trajectory [blue solid]. The dotted lines show the linear and quadratic approximations. The points show the fundamental QNM frequency for perturbation strengths $\epsilon \in \{0.1,\ 1,\ 10,\ 50\}$. 
}
\label{fig:PT-fundamental}
\end{figure}

Figure \ref{fig:PT-fundamental} shows an example of the migration of the fundamental mode in more detail. The plot shows that the linear and quadratic Taylor-series expansions in Eq.~(\ref{eq:series}) [dashed lines] are only accurate for small $\epsilon \lesssim 10^{-2}$, and are poor guides to the global shape of the trajectory. In the limit $\epsilon \rightarrow \infty$, the QNF approaches the zero of $\um(x_0)$. 

\begin{figure}
 \includegraphics[width=8cm]{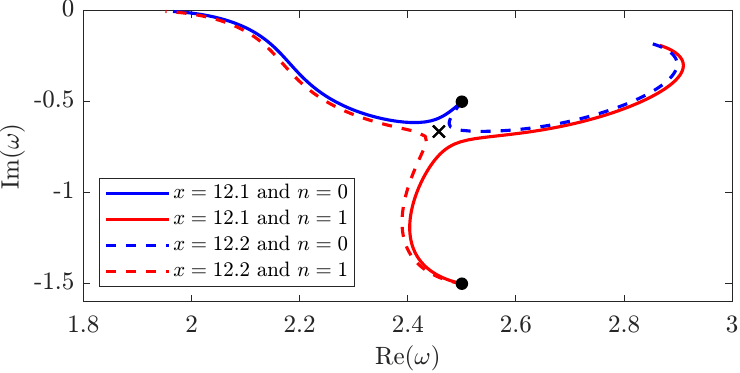}
\caption{
 Examples of the switching of trajectories as the position of the perturbation $x$ is increased. The repelling point is marked with a black cross. The blue and red lines are the trajectories of the $n=0$ and $n=1$ QNFs for $\epsilon \in [0,30]$, with a perturbation at $x_0 = 12.1$ (solid) and $x_0=12.2$ (dashed).
}
\label{fig:PT-switch}
\end{figure}

Figure \ref{fig:PT-switch} shows examples of trajectories that approach a repelling point. In Fig.~\ref{fig:PT-switch}, the $n=0$ and $n=1$ trajectories approach one another near the repelling point; in Fig.~\ref{fig:PT-QNMs} the $n=2$ and $n=3$ modes show a similar behaviour. As the position of the perturbation $x$ is increased, the trajectories `switch' from one attracting point to another. This leads to a global non-smooth change in the individual trajectories. However, the spectrum itself is smoothly deformed with $x$; this is merely an issue of which trajectory is labelled $n=0$ and which is labelled $n=1$. 

\begin{figure}
 \includegraphics[width=12cm]{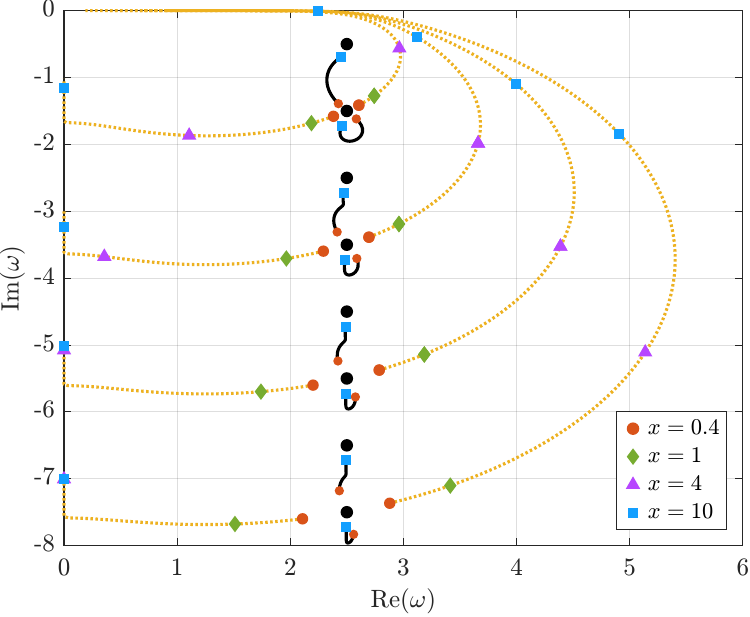} \\
\caption{
 The attracting and repelling points of the QNM spectrum. The attracting (repelling) points migrate along dashed (solid) lines, as the position $x$ of the perturbation is increased from $x = 0.4$ to $x = 90$ (with $\lambda = 5/2$). The open blue circles show the positions of the unperturbed QNFs. % As $x \rightarrow 0^+$, the repelling points approach the odd-$n$ QNFs. As $x \rightarrow \infty$, the repelling points approach the unperturbed QNFs, with the trajectory returning to the same QNF, and one moving to the RP below. For small $x$, the attracting points are on an approximately vertical line to the left of the RP line. For sufficient large $x$ ($x \gtrsim 4$), some attracting points migrate along (but just above) the real axis. This plot is for the Poschl-Teller potential (which peaks at $x=0$) with frequency $\omega = (3/2) \sqrt{27}$. 
}
\label{fig:PT-attractors}
\end{figure}

\subsubsection{Attractors}
In the case where the perturbation is reasonably far from the peak of the potential barrier (i.e.~$e^{-2\hat{x}_0} \ll 1$), we can obtain approximations for the positions of the attracting points, which are determined from the zeros of $\um(x_0)$ and $\up(x_0)$ defined in Eq.~(\ref{eq:uup-pt}). 
Expanding the hypergeometric function at $z=1$, where $z \equiv 1/(1+e^{-2\hat{x}})$ (for $\um$), gives
\beq
\um(\hat{x}) \approx - \frac{i \pi e^{-i \hat{\omega}\hat{x}}}{\sinh(\pi \hat{\omega})} \left[ \frac{e^{-\pi \omega} (z-1)^{-i\omega}}{\Gamma(\tfrac{1}{2} - i \lambda) \Gamma(\tfrac{1}{2} + i \lambda)} - \frac{\Gamma(1-i\omega)}{\Gamma(\tfrac{1}{2} - i (\lambda + \omega)) \Gamma(\tfrac{1}{2} + i (\lambda - \omega)) \Gamma(1 + i \omega)} \right]
\eeq
Hence the w-mode condition $\um(\hat{x}_0) = 0$ corresponds to
\beq
e^{2 i \hat{\omega} \hat{x}_0} \approx \frac{e^{\pi \hat{\omega}} \, \Gamma(\tfrac{1}{2} - i \lambda)}{\Gamma(\tfrac{1}{2} -i \lambda - i \omega)}\frac{\Gamma(\tfrac{1}{2} + i \lambda)}{\Gamma(\tfrac{1}{2} + i \lambda - i \omega)} \frac{\Gamma(1-i\hat{\omega})}{\Gamma(1+i\hat{\omega})}, 
\eeq
which admits $\omega = 0$ as a solution.

On the other hand, for large $x$, expanding the hypergeometric function around $z=0$, with $z \equiv 1/(1+e^{2\hat{x}})$ for $\up$, the first zero of $\up(\hat{x}_0)$ is at
\beq
\omega \approx -i  \left(1 + (\lambda^2 + 1/4) e^{-2 \hat{x}_0} \right) .
\eeq
The $N$th zero is at a frequency $\omega_N$ that take the form
\beq
\omega_N \approx - i \left( N  +  (-1)^{N+1} P_{2N}(\lambda) e^{-2 N \hat{x_0}} \right) ,
\eeq
where $P_{2N}(\lambda)$ is a polynomial of highest power $2N$. 
Hence these attracting points are on the imaginary axis and, for $\hat{x}_0 \gg 1$, they lie close to (but not at) $\omega = -i N$. At $\omega = -i N$, the radial functions $\um(x_0)$, $\up(x_0)$ are divergent, but the ratio $F(\lambda, \omega)$  in Eq.~(\ref{eq:delta-resonance}) remains well-defined. 

%{\color{red} Sentence about repelling point that seem to connect unperturbed QNFs ?}

\subsection{Regge pole spectrum}
\begin{figure}
 \includegraphics[width=12cm]{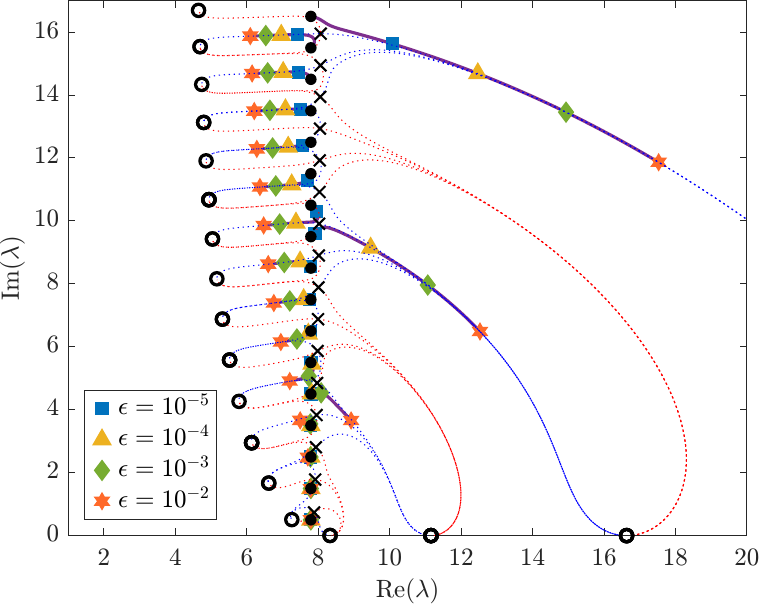}
\caption{
The plot shows the Regge-pole spectrum for a Poschl-Teller potential (\ref{eq:nariai-radial}) with a perturbation at $x_0 = 10 / \sqrt{27}$ of strength $\epsilon$ and frequency $\omega = (3/2) \sqrt{27}$. As $\epsilon$ increases, most RPs move to the left (i.e.~smaller real part), but certain poles ($n=4$, $9$ and $16$) migrate to the right, creating two branches. The attracting (open black circles) and repelling (black crosses) points defined in Sec.~\ref{subsec:flows} are marked. The leftward-moving (rightward-moving) poles migrate towards the attractors with large (small) imaginary part. The blue (red) dotted lines show integral curves of Eq.~(\ref{eq:ode}) that move away from the repellers as $\epsilon$ increases (decreases). As shown, the RP trajectories are attracted towards the blue dotted trajectories.
}
\label{fig:PT-RPs}
\end{figure}

Figure \ref{fig:PT-RPs} shows the Regge pole spectrum under a delta-function perturbation for the P\"oschl-Teller potential, along with the attracting and repelling points, and the integral curves of (\ref{eq:ode}) associated with the latter. The low-$n$ Regge poles are affected least by the perturbation. The higher-$n$ modes are extremely sensitive to small perturbations, and the spectrum splits into two branches (as noted for the Schwarzschild spacetime in Ref.~\cite{Torres:2023nqg}). The left branch is formed by modes that migrate towards attracting points with large imaginary part, and the right branch is formed by modes that migrate towards the attracting points that are close to the real axis. 

Figure \ref{fig:PT-attractors} shows the migration of the attracting and repelling points in the complex plane as the position of the perturbation $x_0$ is modified. The repelling points remain in the vicinity of the unperturbed RPs, whereas the attracting points emerge from the RPs (in the limit $x \rightarrow 0^+$) and then move to the left before progressing along a curve that lies just above the real axis. Once attracting points have moved to the right of the unperturbed RPs (at $x \gtrsim 4$), two branches in the RP spectrum are formed.

\begin{figure}
 \includegraphics[width=12cm]{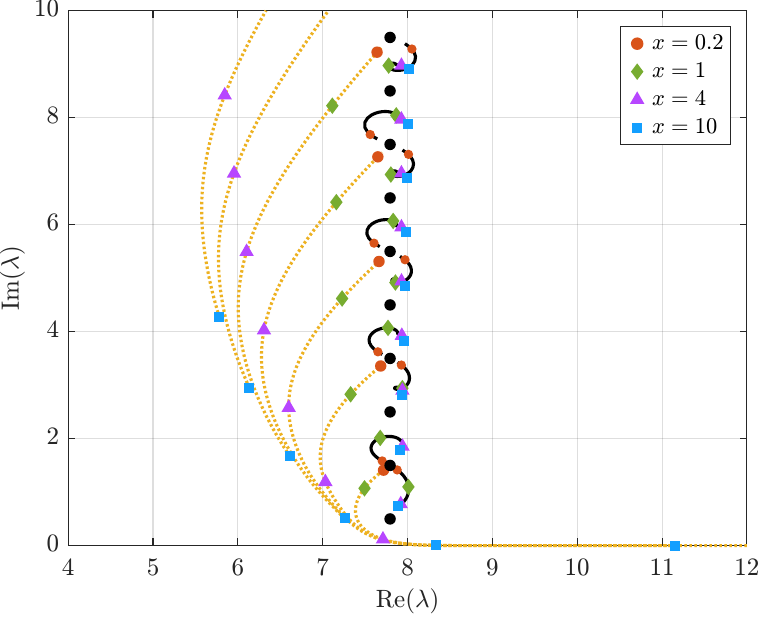} \\
\caption{
 The attracting and repelling points of the Regge pole spectrum. The attracting (repelling) points migrate along yellow dashed (black solid) lines, as the position $x$ of the perturbation is increased from $x= 0.1$ to $x = 10$. The open blue circles show the positions of the unperturbed Regge poles. As $x \rightarrow 0^+$, the repelling points approach the odd-$n$ RPs. As $x \rightarrow \infty$, the repelling points approach RPs, with one trajectory returning to the same RP, and one moving to the RP below. For small $x$, the attracting points are on an approximately vertical line to the left of the RP line. For sufficient large $x$ ($x \gtrsim 4$), some attracting points migrate along (but just above) the real axis. This plot is for the Poschl-Teller potential (which peaks at $x=0$) with frequency $\omega = (3/2) \sqrt{27}$. 
}
\label{fig:PT-attractors}
\end{figure}

 \subsection{Stability and sensitivity}

To quantify the stability of the spectrum, we defined two `threshold' parameters, $\epsilon_{\text{lin}}$ and $\epsilon_{\text{nonlin}}$ in Eqs.~(\ref{eq:epsilon-lin}) and (\ref{eq:epsilon-nonlin}). For perturbation strengths $|\epsilon| \gtrsim \epsilon_{\text{lin}}$ the linear approximation predicts a `large' perturbation in the resonance frequency; for $|\epsilon| \gtrsim \epsilon_{\text{nonlin}}$ the Taylor series expansion itself, Eq.~(\ref{eq:series}), is of questionable validity. 

\begin{figure}
 \includegraphics[width=8cm]{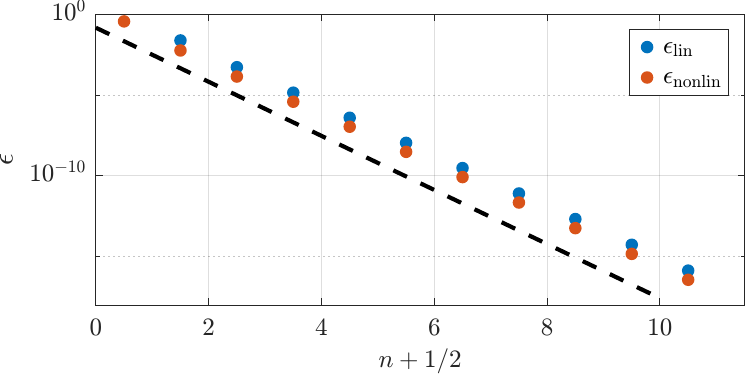} 
 \includegraphics[width=8cm]{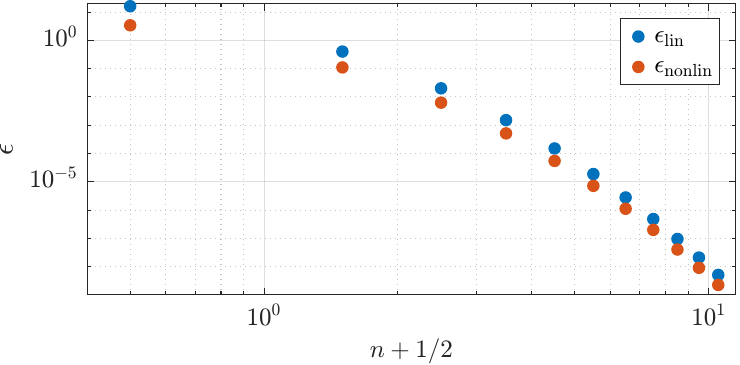} 
\caption{
 Thresholds for linear and non-linear instability ($\epsilon_{\text{lin}}$ and $\epsilon_{\text{nonlin}}$) in overtones  $n=0\ldots 10$ for a perturbation at $x_0 = 10 / \sqrt{27}$ in the Poschl-Teller potential. 
 \emph{Left:} Quasinormal mode stability thresholds with $\lambda = 5/2$. The dashed line shows $\exp(- 2 (n+\tfrac{1}{2}) x_0)$. \emph{Right:} Regge pole stability thresholds with $\omega = \frac{3}{2} \sqrt{27}$. N.B.~the left plot uses a semi-log scale, whereas the right plot uses a log-log scale. 
}
\label{fig:PT-stability}
\end{figure}

Figure \ref{fig:PT-stability} shows $\epsilon_{\text{lin}}$ and $\epsilon_{\text{nonlin}}$ for the QNMs and RPs of the PT potential. These results indicate an exponential sensitivity of higher-overtone QNMs to small perturbations (the Elephant and Flea phenomenon), resulting in the QNM being nonlinearly unstable. The Regge pole overtones are also sensitive, but the thresholds decrease as a power-law rather than an exponential. They appear to be nonlinearly stable in the first few modes but they become unstable for the higher overtones. In the Nariai case, this can be deduced and understood by inspecting the form of the closed-form solutions.

In the QNM case, for $x_0 \gg 1$,
\beq
F'(\omega_n) \approx e^{-2 i \omega_n x_0} \frac{2 \pi (-1)^n \omega_n}{\cosh(\pi \lambda)} \frac{\Gamma(n+1)}{\Gamma(-n - 2 i \lambda)} \frac{\Gamma(-i \omega_n)}{\Gamma(i \omega_n)} .
\eeq
In essence, the leading term is responsible for the the exponential sensitivity of QNFs. Alternatively, the exponential instability of the QNFs for $x_0 \gg 1$ can be understood with reference to the relationship between the linear expansion coefficient and the excitation coefficient in Eq.~(\ref{eq:ef}). 

In the Regge Pole case,
\beq
F'(\lambda_n) \approx \frac{2 i \pi \omega}{\sinh(\pi \omega)} e^{-2 i \omega x_0} \frac{\Gamma(n+1)}{\Gamma(n+1 - 2 i \omega)} \frac{\Gamma(- i \omega)}{\Gamma(i \omega)} .
\eeq
For large overtones, this has the expansion
\beq
F'(\lambda_n) \approx (\lambda_n)^{2 i \omega} \left( 2 i \pi \omega e^{- 2 i \omega x_0} \frac{ e^{\pi \omega}}{\sinh(\pi \omega)} \frac{\Gamma(-i\omega)}{\Gamma(i \omega)} + O(1/\lambda_n^2) \right) .
\eeq
The leading term explains the power-law decay with $n+1/2$ observed in Fig.~\ref{fig:PT-stability}. 

\section{The Schwarzschild spacetime\label{sec:Schwarzschild}}

We now turn our attention and apply our formalism to the astrophysical case where the potential is that of a Schwarzschild black hole, given in Eq.~\eqref{eq:rw}. Unlike in the case of the Nariai spacetime, the closed form solutions of the radial functions $u^{\rm in}_{\lambda\omega}$ and $u^{\rm up}_{\lambda\omega}$ for the Schwarzschild spacetime are only accessible as infinite sums~\cite{Leaver:1986gd,mano1996analytic} which can be difficult to deal with and we use numerical methods to obtain them. This done via a direct integration scheme where the initial conditions for the field are set using a Taylor expansion `close' to the horizon for $u^{\rm in}_{\lambda\omega}$ and at large radius for $u^{\rm in}_{\lambda\omega}$. This poses a signification challenge, particularly when dealing with complex frequencies, since it is known that the QNMs diverge at the horizon and spatial infinity. While it is still possible to obtain numerically accurate QNM with small imaginary part via this direct integration scheme, this procedure is known to fail when looking for higher overtones. To circumvent this issue, other numerical methods have been proposed, such as the standard continued fraction method~\cite{Leaver:1986gd}, to compute the QNFs for large overtone numbers. Unfortunately, our approach requires the actual value of the resonant radial function evaluated at the perturbation location, hence when dealing with complex frequencies, we will be limited to studying the fundamental QNM. However, this limitation does not appear when dealing with RPs, since in this case the frequencies are real and the resonant modes are oscillatory at the horizon and infinity. This means that we can accurately compute the migration for the RPs spectrum, including the higher overtones.

\textbf{Migration of resonances --}
We compute the migration of the fundamental QNFs and RPVs from the dynamical equations Eqs.~\eqref{eq:ode} as the strength of the perturbation, $\epsilon$, is increased. We have also computed the trajectories using an extension of the standard continuous fraction scheme~\cite{Leaver:1986gd}, as described in \cite{Benhar:1998au,OuldElHadj:2019kji,Torres:2022fyf,Torres:2023nqg}, and checked that the two methods lead to the same results.

\begin{figure}[!h]
 \includegraphics[width=8cm]{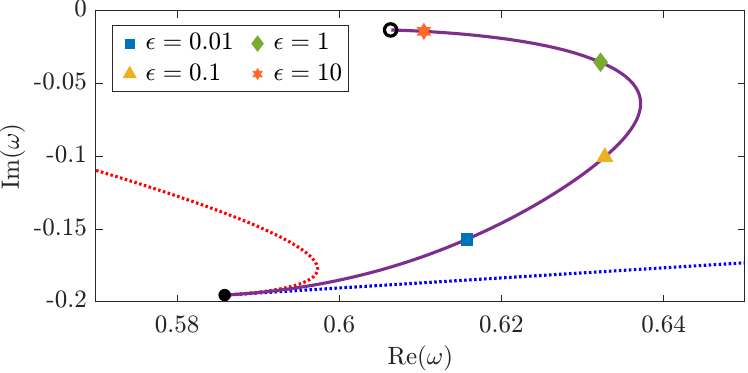} 
 \includegraphics[width=8cm]{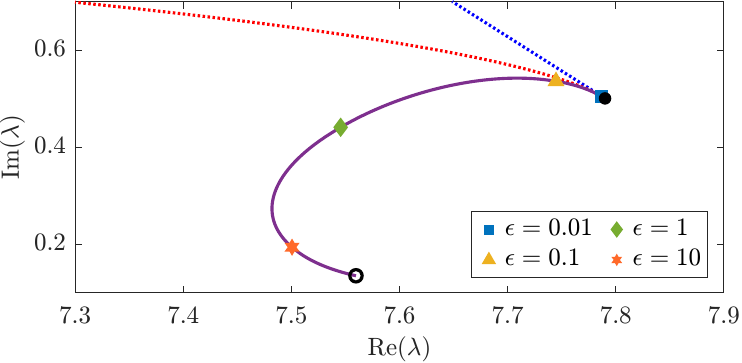} 
\caption{
 Migration of the fundamental QNF for $\ell=1$ (left) and RP for $\omega = 3$ (right) in the perturbed Schwarzschild potential as the strength of the perturbation $\epsilon$ is increased from $0$ to $500$. The perturbation is location at $x_0 = 10$.
}
\label{fig:QNM_migration_fundamental}
\end{figure}

Figure \ref{fig:QNM_migration_fundamental} shows the migration of the fundamental QNF and RPV in the complex plane as the strength of the perturbation is increased. We can see that the resonances migrate from their unperturbed value (blue circle) towards the attractor (black circle). One striking difference between the behaviour of the QNF and the RPV is the rate at which the linear and quadratic approximations (dashed lines), given in Eq.~(\ref{eq:series}) break down. For the QNFs, these approximation break down almost instantly, $\epsilon <10^{-2}$ while the quadratic approximation agrees up to $\epsilon \approx 10^{-1}$ for the RPV.

\begin{figure}[!h]
 \includegraphics[width=12cm]{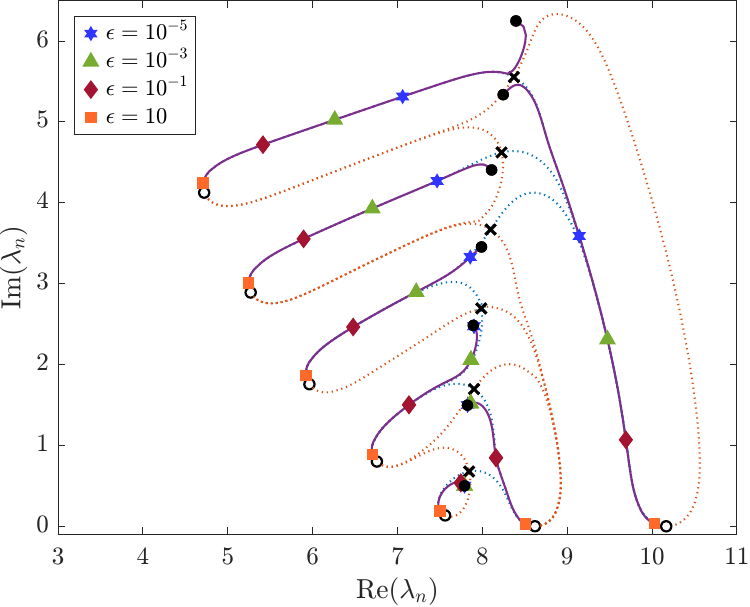}
\caption{
The plot shows the Regge-pole spectrum for the Schwarzchild BH (\ref{eq:rw}) with a perturbation at $x_0 = 10 $ of strength $\epsilon$ and frequency $\omega = 3$. The migration plot is very similar to the Nariai spacetime, see Fig.~\ref{fig:PT-RPs}. As $\epsilon$ increases, most RPs move to the left (i.e.~smaller real part), but certain poles ($n=1$ and $5$) migrate to the right, creating two branches. The attracting (empty black circles) and repelling (black crosses) points defined in Sec.~\ref{subsec:flows} are marked. The leftward-moving (rightward-moving) poles migrate towards the attractors with large (small) imaginary part. The blue (red) dotted lines show integral curves of Eq.~(\ref{eq:ode}) that move away from the repellers as $\epsilon$ increases (decreases). As shown, the RP trajectories are attracted towards the blue dotted trajectories. 
}
\label{fig:Schw-RPs}
\end{figure}
Figure \ref{fig:Schw-RPs} shows the Regge pole spectrum on Schwarzschild spacetime under perturbation. Qualitatively, there are  similarities with the spectrum of the Nariai spacetime shown in Fig.~\ref{fig:PT-RPs}. In particular, we see that the attractors arrange themselves in two branches: one lying close the real axis, corresponding to quasi-bound states trapped between the light-ring and the infinite delta function perturbation, and second one with comparable imaginary part than the unperturbed spectrum. The RPVs migrate from their unperturbed value towards both branches, with their trajectories governed by the location of the repelling points (black crosses). We can clearly see that the higher overtones are a lot more sensitive to the perturbation.

\section{Discussion and conclusions\label{sec:conclusions}}
In this work we have investigated the deformation and stability of black hole resonance spectra under localised perturbations of the effective potential. By treating QNMs and RPs on equal footing as resonances defined through global boundary conditions and using a dynamical system perspective, we have clarified how small local perturbations can lead to large modification of the spectrum in the complex frequency/angular momentum plane.

A central result of this study is the identification of resonance migration as a dynamical flow in the complex plane. Rather than appearing or disappearing abruptly, resonances move continuously along well-defined trajectories as the perturbation strength is varied. These trajectories are governed by a small number of attracting and repelling points whose locations are fixed by the analytic structure of the unperturbed system and by the position of the perturbation. 
A deeper understanding of these trajectories, both as isolated objects and as a collective system~\cite{motohashi2025resonant,macedo2025exceptional} may provide valuable insights in understanding the apparent tension between the observed unstable resonance spectra and the stability of associated observables (see also~\cite{lenzi2025korteweg} for an alternative approach). Indeed, it has been shown that many observable quantities constructed from the resonances, such as the scattering cross-section~\cite{Torres:2023nqg,syu2025regge} or the greybody factors can exhibit stability or instability under potential perturbations, depending on the frequency and angular momentum~\cite{oshita2023thermal,oshita2024greybody,okabayashi2024greybody,oshita2024stability,rosato2024ringdown}.

In addition, our work identifies and separates different types of instability. Through a perturbative approach, we find relevant to introduce the notions of linear, nonlinear and anomalous instability.
By linking the coefficient of the linear perturbation expansion to the physical QNM excitation factor, we have provided an alternative demonstration of the “Elephant and Flea” phenomenon and an explanation for its origin. Furthermore, this distinction provides characteristic scales for the perturbation linked to the spectral instability effect, which is known to be strongly dependent on the choice of norm~\cite{Gasperin:2021kfv}.

Using idealised delta-function perturbations and the Nariai spacetime, we have isolated the essential mechanism responsible for instability and allow for exact or semi-analytic control over the spectrum. When applied to the Schwarzschild spacetime, the same qualitative behaviour persists. Both QNFs and RPVs exhibit strong sensitivity to localised perturbations, with migration trajectories organised by a similar attractor–repeller structure identified in the Nariai spacetime. Importantly, the perturbations considered here can be placed arbitrarily far from the light ring, reinforcing the conclusion that spectral instability is not controlled solely by local photon-orbit physics, but instead reflects the global nature of resonance boundary conditions. 

While we have focused on delta-function perturbations for their analytical tractability, the dynamical-systems framework we introduced is not limited to this case. Indeed, the resonance migration as a flow in the complex plane, governed by attracting and repelling points is expected to persist for more general localised perturbation. This is supported by
several observations of the spectral instability phenomenon (in the form of the ''Elephant and Flea" effect) in studies using smooth perturbations~\cite{Barausse:2014tra,Cheung:2021bol,Torres:2023nqg}. This suggests that the instability arises from the global nature of the resonance boundary conditions and the sensitivity of the Wronskian condition \eqref{eq:Wronsk1} to local changes in the potential. We also note that since Eq.~\eqref{eq:Wronsk1} stems from a perturbative analysis, generic local perturbations of compact support will lead to a spectrum migration essentially identical to the one produced by the delta-function perturbation for $0<\epsilon\ll1$.
Furthermore, the use of the delta-function perturbation allows us to study the migration of resonances beyond the small $\epsilon$ regime, i.e. when the added local structure is of similar size than the potential. This situation has been extensively explored~\cite{Leung:1997was,Leung:1999iq,Barausse:2014tra,hui2019quasinormal} and leads to the emergence of echoes~\cite{cardoso2017observational}. The regime $\epsilon \gg1$ resembles an effective hard wall and we have described the resonances in this regime as remnants of the original unperturbed resonances.

Finally, the dynamical-systems perspective introduced in this work may prove useful beyond black hole physics. For example, in the study of resonances in analogue gravity systems to explain the origin of resonances in semi-open systems~\cite{solidoro2024quasinormal} or in photonics where QNMs are routinely used to design nanoresonators~\cite{wu2024designing}.

\begin{acknowledgments}
%\textit{Acknowledgements.---} 
TT gratefully acknowledges Antonin Coutant and José Luis Jaramillo for insightful discussions and their interest in this project. While the results presented here were obtained some time ago, their shared enthusiasm played a important role in the completion of the present manuscript.
S.D. acknowledges financial support from the Science and Technology Facilities Council (STFC) under Grants No. ST/X000621/1 and No. ST/W006294/1.
\end{acknowledgments}

\appendix

\section{Regular perturbation theory\label{app:perturbation-theory}}
In this section we seek the radial function $\uin(x)$ that satisfies Eq.~(\ref{eq:radial-equation}) for a perturbation $\delta V(x)$ with support on $x \in [a,b]$ only. First, Eq.~(\ref{eq:radial-equation}) is recast in the form
\beq
\left[ \frac{d^2}{dx^2} + V_{\lambda \omega}(x) \right] u(x) = - \epsilon \delta V(x) u(x).
\eeq
Next, assume that $u(x)$ admits a regular power series expansion of the form
\beq
u(x) = u_0(x) + \epsilon u_1(x) + \epsilon^2 u_2(x) + \ldots 
\eeq
At zeroth order, $u_0(x) = \um(x)$. At first order,
\beq
\left[ \frac{d^2}{dx^2} + V_{\lambda \omega}(x) \right] u_1(x) = -  \delta V(x) u_0(x).
\eeq
This ODE has the formal solution
\beq
u_1(x) = - \int g(x,x') \delta V(x') u_0(x') dx' ,
\eeq
with the Green's function  
\beq
g(x,x') = \frac{1}{W_0} \left(\Theta(x-x') \up(x) \um(x') + \Theta(x'-x) \um(x) \up(x') \right) ,
\eeq
where $W_0$ is defined in Eq.~(\ref{eq:W0}). Consequently, the first-order function $u_1$ on either side of the region of compact support is
\begin{align}
u_1(x < a) &= u^{\text{homog.}}(x) - \frac{\um(x) c^-_{\lambda \omega}}{W_0} , \\
u_1(x > b) &= u^{\text{homog.}}(x) - \frac{\up(x) c^+_{\lambda \omega}}{W_0} ,
\end{align}
where $u^{\text{homog.}}(x)$ is a homogeneous solution of the unperturbed equation (i.e.~a linear sum of $\up$ and $\um$), and the coefficients $c^{\pm}_{\lambda \omega}$ are defined in Eq.~(\ref{eq:cco}). To ensure the left-hand `in' boundary condition is satisfied, we set $u_1(x < a) = 0$ to fix the homogeneous piece uniquely, leading directly to Eq.~(\ref{eq:u-general}).

\bibliographystyle{apsrev4-1}
\bibliography{main}

\end{document}